\documentclass[12pt]{iopart}

\usepackage{graphicx}
\usepackage{booktabs}
\usepackage{numprint}
\usepackage{appendix}
\usepackage{xcolor}

\begin{document}

\title [Statistical properties of multimodal mobility]{A survival model to explain the statistical properties of multimodal mobility}

\author{C. Mizzi$^{\ast}$, A. Fabbri$^{\ast}$, G. Colombini$^{\ast}$, F. Bertini$^{\dagger}$, A. Bazzani$^{\ast}$}

\address{$\ast$Laboratory of Complex Systems Physics\\
Physics and Astronomy Department- University of Bologna\\ 
INFN - sezione di Bologna\\
$\dagger$ Mathematical, Physical and Computer Sciences Department, University of Parma}
\ead{armando.bazzani@unibo.it}
\vspace{10pt}
\begin{indented}
\item[]December 2021
\end{indented}

\begin{abstract}
The statistical properties of human mobility have been studied in the framework of Complex Systems Physics. Taking advantage from the new datasets made available by the information and communication technologies, the distributions of mobility path lengths and of trip duration have been considered to discover the fingerprints of complexity characters, but the role of the different transportation means on the 
 statistical properties of urban mobility has not been studied in deep. In this paper we cope with the problem of the existence of universal features for pedestrian, bike and vehicular urban mobility. In particular, we propose the use of travel time as universal energy for the mobility and we define a simple survival model that explains the travel time distribution of the different mobility types. The analysis is performed in the metropolitan area of Bologna (Italy), where GPS datasets were available on individual trips using different transport means. Our results could be helpful for the realization a multimodal sustainable mobility in the future cities, compatibly with the citizens propensities to use the different transport means.
\end{abstract}

%
\noindent{\it Keywords}: GPS data, travel time distribution, utility function, survival models
%
%
%
%
\section{Introduction}

Human mobility is a consolidated research field for Complex Systems Physics that  have been considered in many previous papers to study the complexity nature of human mobility\cite{brokmann2006,gonzalez2008,song2010}, to build models that relate the microscopic individual behaviour to the macroscopic properties of the empirical statistical distributions\cite{bazzani2007,bazzani2010,yanquing2011,wang2014} and to the develop a Statistical Physics for cognitive systems\cite{gallotti2012,noulas2012}. The information communication technologies allowed to get data on individual mobility\cite{giannotti2011,liang2012,gallotti2015}. The relevance for applications is related to the development of sustainable urban mobility, the improvement of the life quality and the realization of the smart cities\cite{batty2012}. One of the key issue towards a sustainable mobility is to reduce the use of the private transportation means in the cities and to improve a multimodal mobility\cite{tran2021}. A possible approach requires the understanding how the individuals realize the mobility demand in a city when different mobility networks are available.
We refer to the possibility of using different transport means taking into account both the individual decision mechanisms and the physical interactions in the mobility networks\cite{bathelemy2016,ding2018}.
The decision to use different transport means could depend on the length of planned trip, the expected duration and the perceived convenience of the choice\cite{akiva1999}. For example, in the case of cycling or pedestrian mobility, the trip length is directly related to the fatigue necessary to perform the trip, whereas in case of car mobility one has to consider the stress related to the driving in traffic conditions, the accessibility of the city to private cars and the availability of parking places. The planning of individual daily mobility can also influence the choice of the transport means due to the necessity of performing several activities during the same day.  A Statistical Mechanics approach could point out some universal features of the empirical distribution functions of human mobility and could suggest observables able to explain how individuals use different transport means\cite{kolbl2003}. In this work we take advantage from the availability of two datasets in the metropolitan area of Bologna (Northern Italy) to study statistical properties of multimodal mobility. These datasets contain GPS data on individual paths performed using different transport means and they provide accurate information on the path length, the travel time and the average velocity of each path on a population sample. In the paper we perform a detailed analysis of the statistical features of these observables for the different transport means to show the consistency with analogous results obtained using other datasets\cite{bazzani2010,liang2013} and the possibility of defining a \textit{mobility energy}\cite{kolbl2003,gallotti2012}. Our results suggest that the travel time can be considered as an universal mobility energy consistently with the concept of travel time budget proposed in the literature\cite{marchetti1994}.
Then we consider the possibility of using a simple model that explains the main features of the travel time distribution for the different transport means and that could be related to the decision mechanism for the choice of them. Our approach is inspired by concept of cost function and logit models introduced by the economists to model decision mechanisms\cite{akiva1999}. Making a correspondence between  the mobility cost function and the energy concept in Statistical Physics, we propose a survival model\cite{gallotti2015} able to reproduce the empirical travel time distributions in the cases of pedestrian, bike and private car mobility using three parameters whose physical meaning is the time 
cost, the convenience and the typical trip duration. The three parameters explain the differences in the travel time distributions and the observed collapse of the distributions when one normalizes the travel times with respect to the average value. The three time scales could be useful to compare the features of urban mobility in a city where specific policies are realized to reduce the private traffic. The proposed survival model could be also used for the development of urban mobility models that introduce a decision mechanism to simulate the urban mobility using different transport means.\par\noindent
The paper is organized as follows: in the second section we illustrate the
quality of the datasets used for the analysis (more details are reported in the Appendix), in the third section we define a survival model for multimodal urban mobility and in the fourth section we show the
results of our analysis. Finally, we draw conclusions and discuss some perspectives.

\section{The mobility datasets}

\begin{figure}[tbp]
  \begin{center}
    \includegraphics[width=0.765\textwidth]{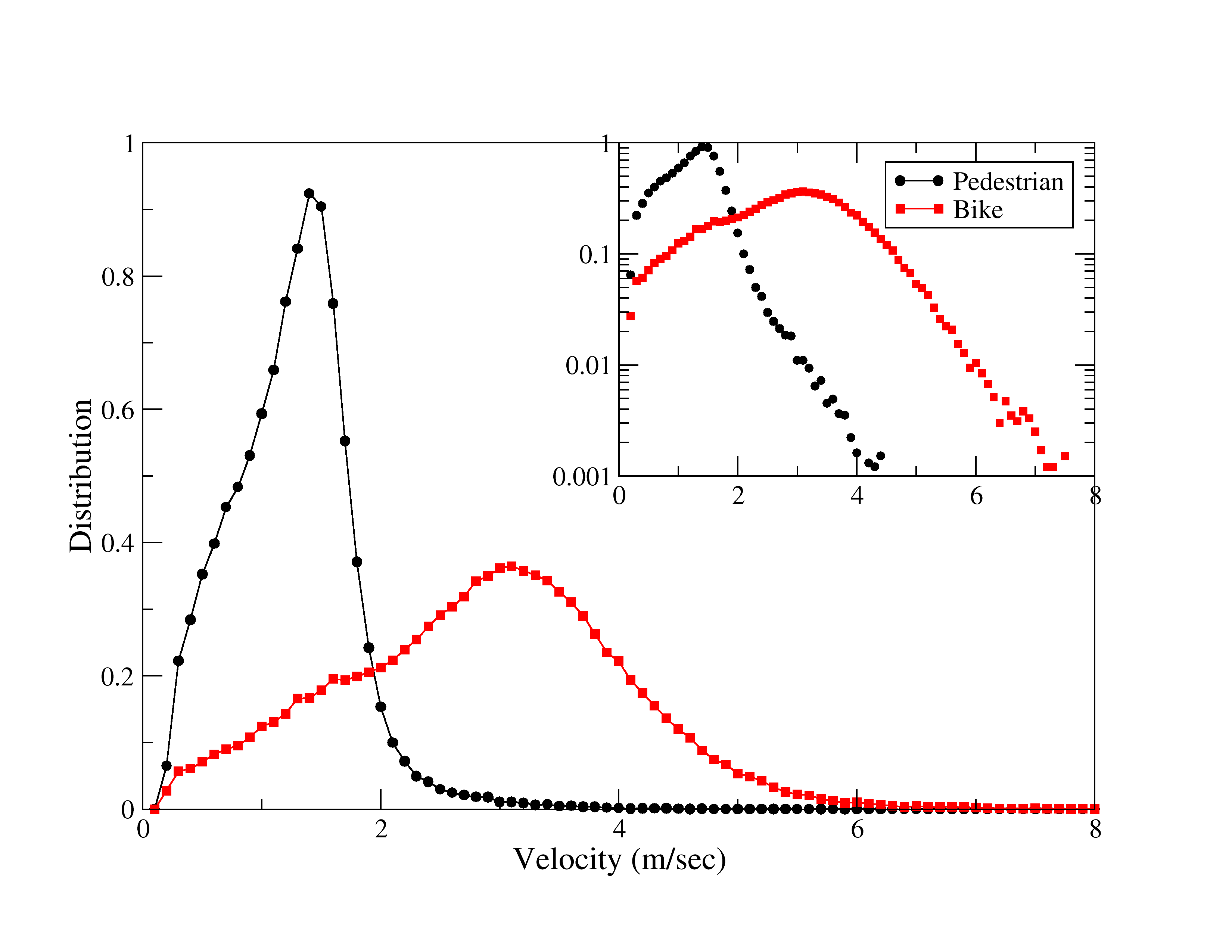}
    \includegraphics[width=0.765\textwidth]{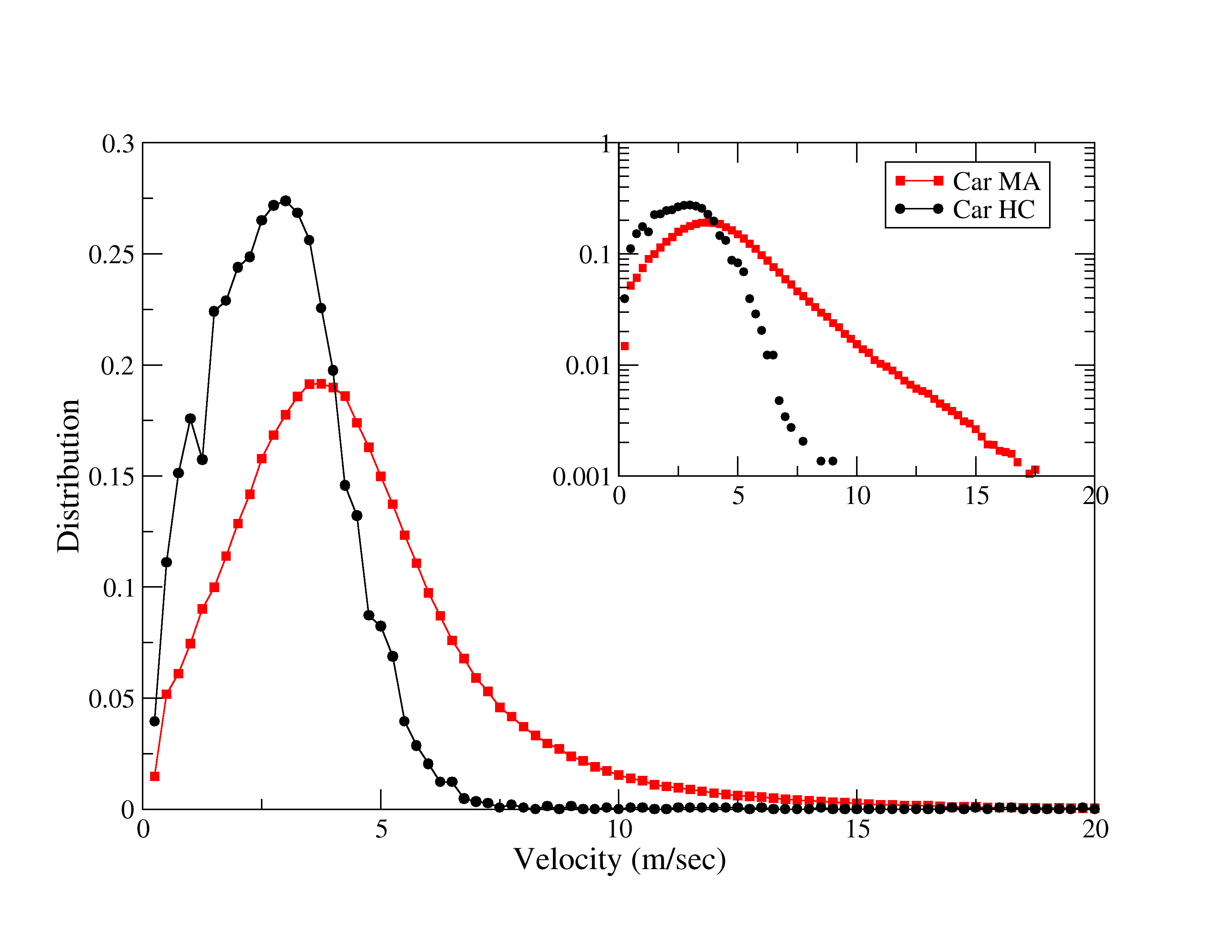}
  \end{center}
  \caption{Top picture: average velocity distribution for the pedestrian paths (black circles) and of the bike path (red squares) computed using the BM dataset. The inset shows the same distribution in a semilog
  scale to highlight the different decaying behaviour.  
  Bottom picture: distribution of the average velocity of the car paths recorded in the OT dataset in the whole metropolitan area: the black circles refer to the car trips in the historical center (HC) during rush hours, whereas the red squares refer to the car trips in the whole metropolitan area (MA). The inset shows the same distributions in a semilog scale.
  We remark the exponential decaying of the velocity distribution for the car paths in the MA.}
  \label{fig2}
\end{figure}

To study the statistical properties of multimodal mobility we take advantage from the availability of different datasets on urban traffic that contain a sample of anonymized trajectories in the metropolitan area of Bologna: a city in the North of Italy with $\simeq$\numprint{400000} inhabitants and a large historical center that induces great mobility demand from peripheral areas.  
From one hand we had access to the Bella Mossa (BM) dataset that provides information on the bike and pedestrian mobility recorded during on a period of 6 months (from April to September 2017)
through a specific app offered to the citizens. From the other hand, we used the Octo Telematics (OT) dataset that contains GPS data on the vehicle trajectories, recorded by insurance reasons during the month of September 2016. Both datasets allow to reconstruct the individual mobility paths in the metropolitan area of Bologna (see. the
\ref{app:data} for more details). They are not open datasets and only aggregated information can be shared, but they have been accessible following scientific collaboration agreements. 
The GPS data quality was checked by using a georeferencing procedure on the road network derived from the OpenStreetMap project \cite{OpenStreetMap} and the consistency between the expected velocity of a trajectory associated to a transport means and the instant velocity computed directly from the
data. The georeferencing procedure select the roads according to a nearness algorithm (with a tolerance $\le 50$ m) and a consistency request with the other data of the same path. 
A detailed description of the data set quality is reported in the \ref{app:data} where we briefly discuss the procedure to reconstruct the single mobility paths \cite{bazzani2010,gallotti2012}.
In fig. \ref{fig2} the velocity distributions for pedestrian, bike and car trips are shown. The velocity distributions for the pedestrian and bike mobility (top figure) indicate that the information in the dataset are in general correct and we have studied the statistical properties of the recorded mobility according to this assumption. We remark the different features of the distribution: in the first case we have an average velocity of $1.19$ m/sec (with a peak $\simeq 1.4$ m/sec) and a sharp decay of the distribution for velocities higher than $2$ m/sec due to the physical limit of the pedestrian velocity, whereas the variability of low velocities is due to the many stops that may occur during the trip. In the bike case we have an average velocity of $3.05$~m/sec which is also the mode of the distribution but the dispersion of the distribution is larger with an exponential-like decaying for greater velocities (see the inset in the figure).
The existence of paths with very low average bike velocity could be the consequence of the traffic rules or of short stops. In the fig. \ref{fig2} (bottom) we plot the average velocity distribution for the
car trips distinguishing between the in the whole Bologna metropolitan area (MA) where a ring road is present for fast moving, and the trips in the historical center (HC) during rush hours where traffic restriction rules are applied (see \ref{app:data}). 
The velocity distribution computed using only the car trips in the HC during rush hours (7:30-8:30 of a working day using a statistics of \numprint{5600} trajectories)
has bell shape centered at the average value $2.84$~m/sec ($\simeq 10.2$~km/h) and sharply decreasing. This behaviour suggests the presence of a strong interaction among vehicles and strict speed limits in the road network. On the contrary the velocity distribution in the MA is characterized by an average velocity of $4.53$~m/sec ($16.3$ km/h) with a lower probability
to observe slow trips and an exponential decaying as highlighted by the inset in a semilog scale. 
The velocity distribution for the HC trips is probably the result of the traffic rules in the HC road networks where many crosses are present, whereas
the car velocity distributions in the MA should be interpreted as the result of the coexistence of different type of road networks in the same area  characterized by a different travel velocity even if some features are universal. The urban road network near the HC is characterized by the frequent crossing points, whereas the country road network has larger streets, that connect the peripheral area to the center 
such as the highway and the Bologna ring road. A possible explanation for the car velocity distribution
is reported in the \ref{app:data}.
To understand the features of each mobility type considered in the paper , we report in the table \ref{table1} the average values of the velocity, path length and travel time distributions.
\begin{table}[b!]
  \begin{center}
    \begin{tabular}{l|c|c|r}
      \toprule
      Transport means & Velocity $V_m$ & Path length $L_m$ & Travel time $T_m$ \\
      \midrule
      pedestrians & 1.19 m/sec  & 1.81 km & 31.6 min. \\
      bikes & 3.05 m/sec  & 3.49 km &  24.5 min. \\
      cars HC & 2.84 m/sec  & 2.01 km  &  11.3 min. \\
      cars MA & 4.53 m/sec  & 3.91 km & 13.4 min. \\
      \bottomrule
    \end{tabular}
  \end{center}
  \caption{Average values for the different mobility distributions: the statistical error on the parameter values is implied in the last digit.}
  \label{table1}
\end{table}
The average velocity is a characteristic of the pedestrian and bike mobility, whereas it depends on the traffic conditions and the road network for cars. In the considered cases the average velocities are much smaller than the typical limit velocity for a urban road network (i.e. $50$~km/h). In particular the average velocity in the HC during rush hours is also smaller than the bike velocity. We also observe that the average path length for MA cars ($3.91$~km) is near of the average bike path length $3.49$~km, whereas the average path length for HC car $2.01$~km is near the pedestrian path length $1.81$~km. 
In particular, the use of car in the HC is perceived convenient also for small trips for the possibility of performing a complex mobility, for which the public transportation would required much more time.
It is highly probable that the citizen sample in the BM dataset has a bias with a propensity to perform the pedestrian and bike mobility. Indeed, in the dataset are present long pedestrian and bike paths, that are probably due to trained individual and increase the average path length values.
Moreover, if one computes the expected travel times from the path length and the velocity, the corresponding values ($L_m/V_m=25.3$ min. 
for pedestrians and $L_m/V_m=19.0$ min. for bikes) are shorter than the average travel time $T_m$ computed using the single trips.
This could be understood since some people may have forgotten to switch off the Bella Mossa app at the end of a trip, so we choose to use these values as $T_m$ in the sequel. 
However, the data suggest as the cars are considered for private mobility even for short trips probably because the pedestrian mobility is considered not convenient or too tiring, whereas the bike is an alternative transport mean to the car since it realizes a similar mobility. The choice of bike mobility could to be related to energy required by riding a bike or to the discomfort to share the road network with cars.

\subsection{Statistical properties of multimodal mobility}

\begin{figure}[tbp]
  \begin{center}
 \includegraphics[width=0.765\textwidth]{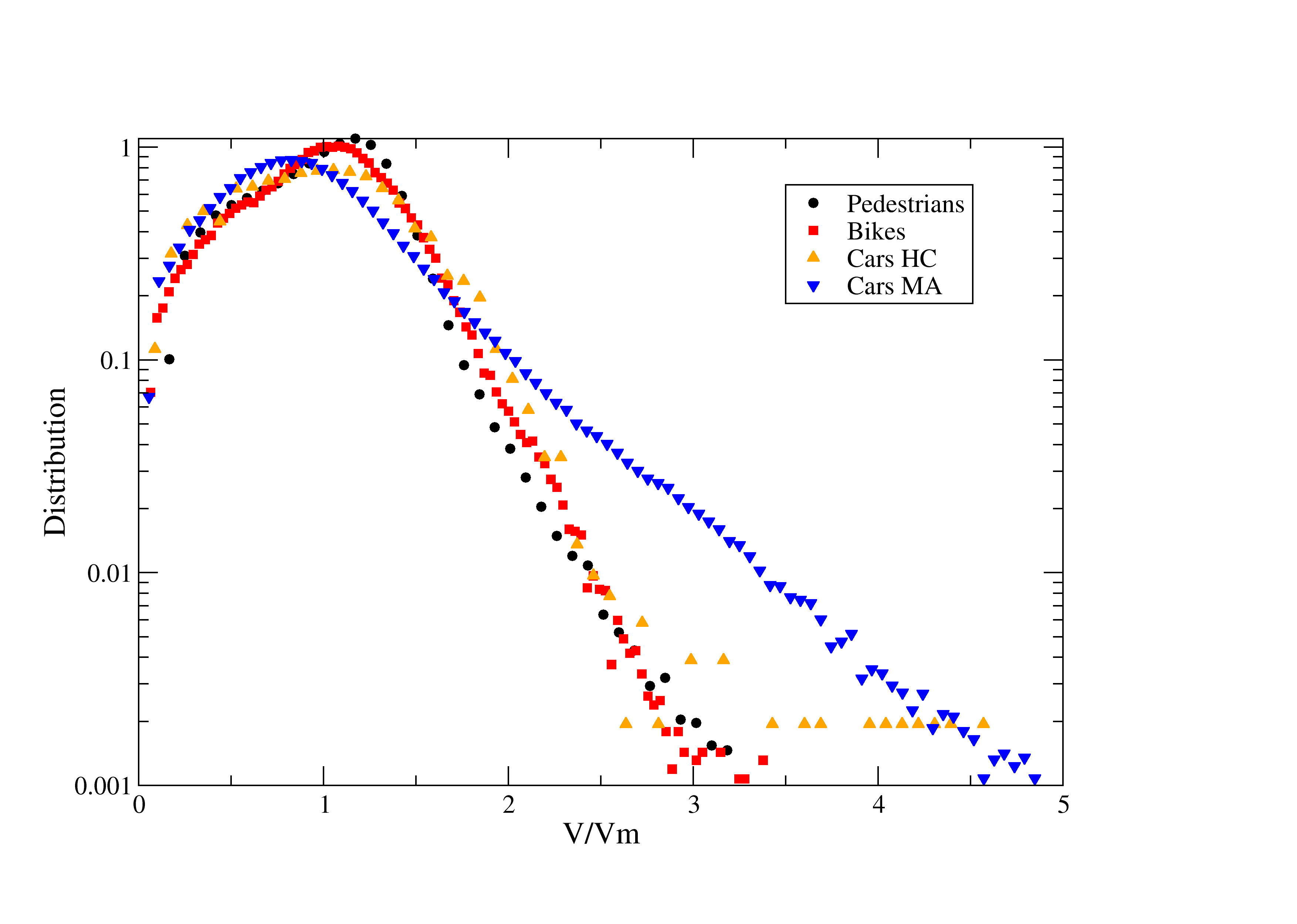}
 \end{center}
  \caption{Distributions of the normalized average velocity for all the considered datasets using a semilog scale: in the inset we give the correspondence of the different symbols.  We remark the  collapse of the distributions for the pedestrian, bike and car HC mobility, whereas the distribution for the car mobility in the Bologna MA shows a different behaviour.}
  \label{fig3}
\end{figure}

To understand the statistical properties of the mobility performed using the different transport means we study the behaviour of the distribution functions of average velocity, path lengths and travel time. 
The velocity distribution of the different transport means is mainly characterized by its average value, but we consider the problem of the existence of universal features
of the different mobilities related to the shape of the distributions.
If we compare the normalized average velocity distributions for the different mobility types, we observe a collapse for the pedestrian, bike and car HC distributions, whereas the car distribution in the MA shows a different decaying pattern for high values $V/V_m$ (see fig. \ref{fig3}). 
This behaviour could depend on the features of the mobility network (e.g. the presence of several crossings) as well as on the individual behaviour (e.g. the individual heterogeneity implies different velocities). Indeed, the pedestrian and bike mobility share the same road network with the car HC mobility in most paths and the effect of frequent stops introduces an heterogeneity in the car dynamics similar to the individual heterogeneity in walking and riding a bike.
On the contrary, the country roads network and the highway contribute mainly to the car MA mobility and the underlying road network can be viewed as a multilayer network with different travel velocities for the different layers, on which the individuals perform their mobility using strategies to move in the different layers\cite{gallotti2016}. Individuals have the possibility to improve the mobility efficiency reducing the travel time for long trips (see also fig. \ref{fig7} and the \ref{app:data} for a possible explanation).\par
\begin{figure}[tbp]
  \begin{center}
    \includegraphics[width=0.765\textwidth]{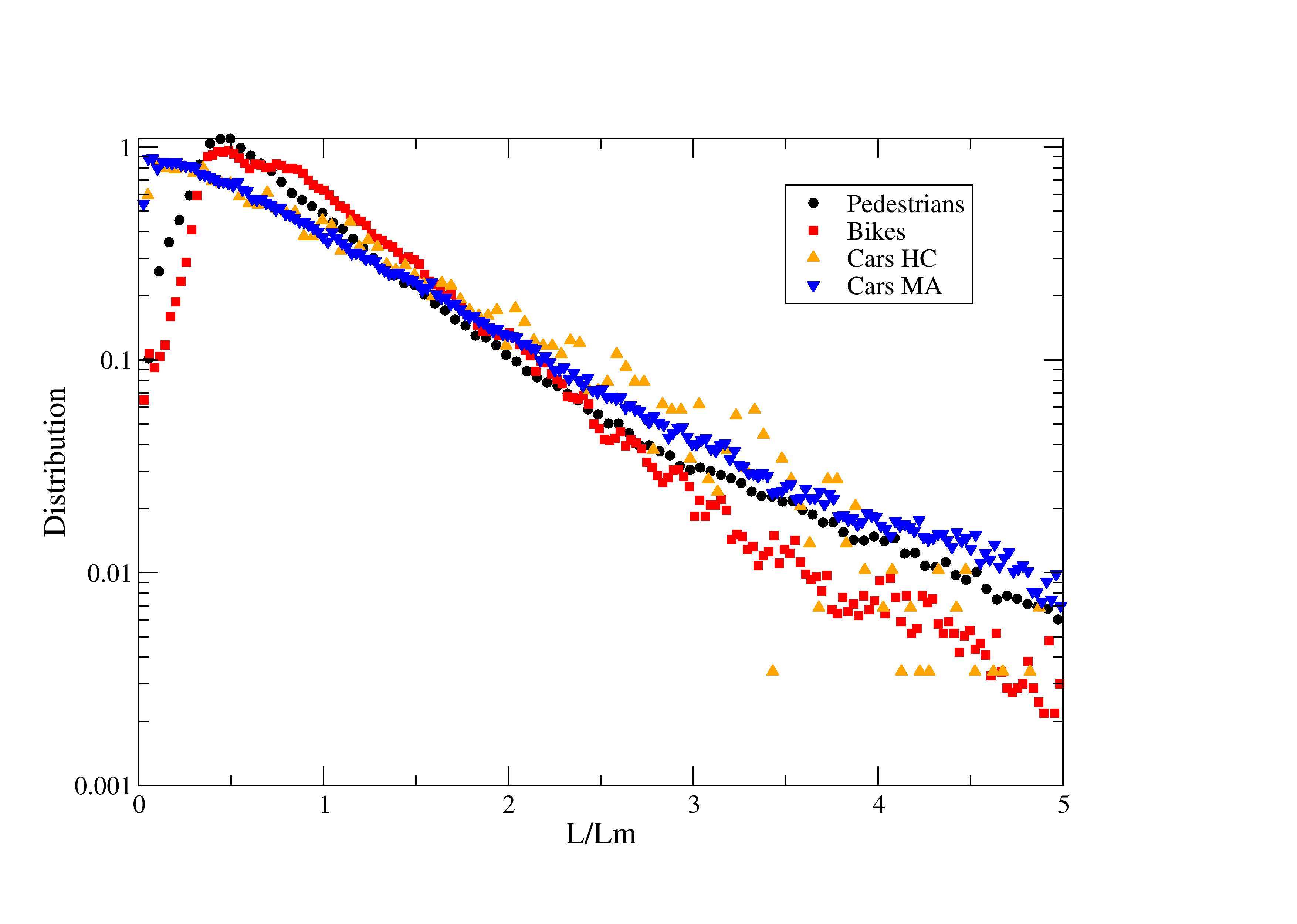}
    \includegraphics[width=0.765\textwidth]{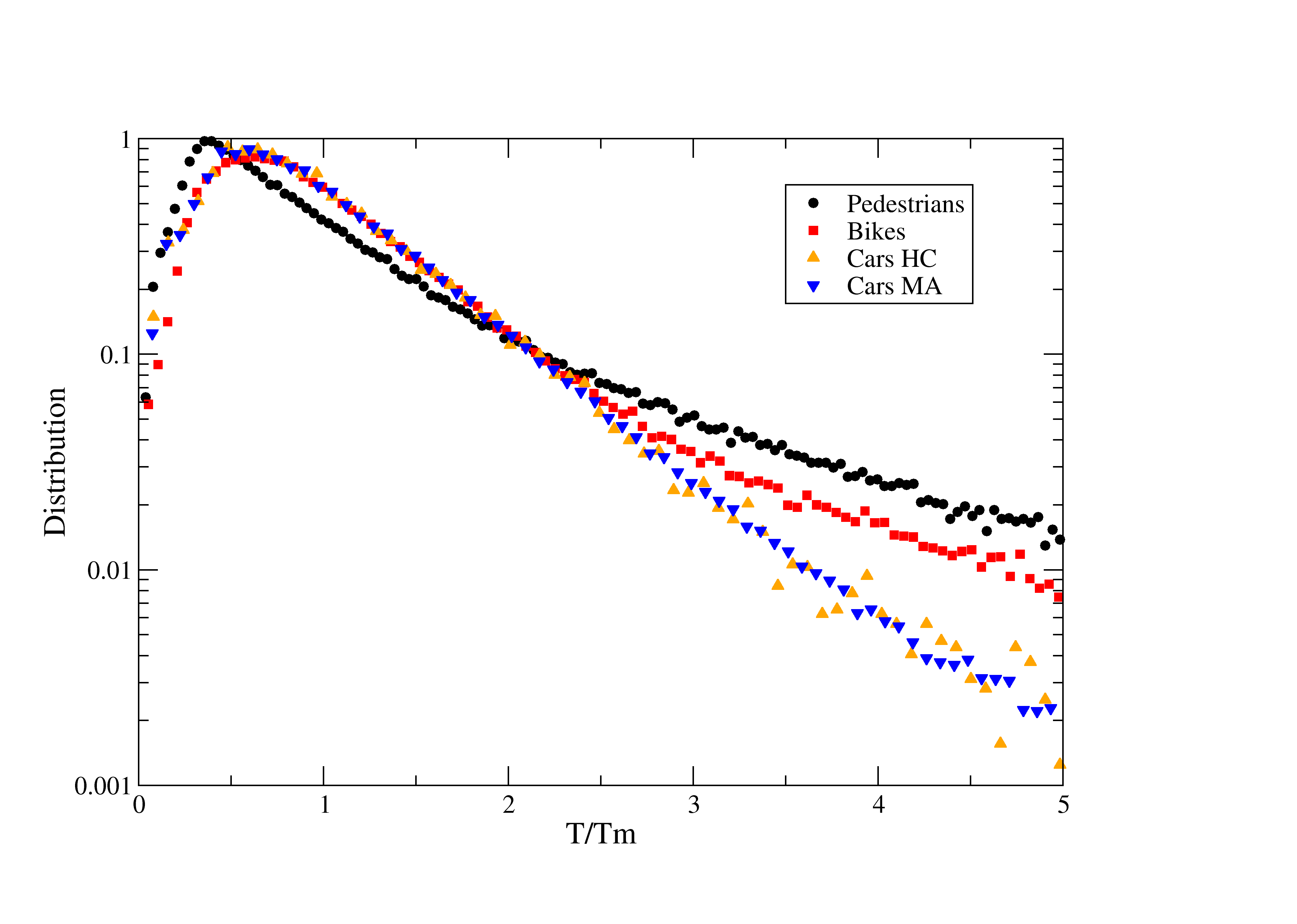}
  \end{center}
  \caption{Top picture: normalized average path length distributions for all the considered datasets: pedestrians (circles), bikes (squares) cars HC (triangles up) and cars MA (triangles down) using a
  semilog scale. Bottom picture: normalized average travel time distributions for all the considered datasets using the same convention for the symbols. It is noteworthy to observe as
  the distributions for the bike and car mobility tend to collapse.}
  \label{fig5}
\end{figure}

\noindent
The path length distribution is not affected by the previous arguments, but it plays a fundamental role to understand which mobility is realized by the different transport means. Previous results\cite{bazzani2010,liang2013} have pointed out as the urban mobility is mainly characterized by an exponential-like decaying in the path length distribution since it reflects the habit of people to perform a local mobility. From a Statistical Physics point of view, the exponential decaying can be explained by using a Maximum Entropy Principle and introducing the concept of \textit{mobility energy}\cite{kolbl2003}. The path length distributions for the different mobility types are characterized by the mean values reported in table \ref{table1} and in fig. \ref{fig5} we show the normalized path length distributions to highlight the universal features. The semilog scale suggests that all the normalized path length distributions are characterized by an exponential-like decaying but it is also possible to observe some structures.
The pedestrian and the bike distributions show an initial growth corresponding to the short paths followed by a fast exponential decrease. But the pedestrian distribution has a slope change at $L/L_m\simeq 2.5$ that could denote the presence of long paths due to sport activities, whereas the bike distribution shows a bimodal structure that could be related to the presence of two types of mobility demand (i.e. one could distinguish the bike mobility in the historical center from the bike mobility in the peripheral areas).
The car mobility provides two distributions with a slower exponential decaying, where the short paths $L\ll L_m$ do not affect the initial behaviour. These distributions would imply that private cars are used to perform a complex mobility that realizes many activities (i.e. not a simple origin-destination mobility).
Finally, we have compared the normalized travel time distributions since mobility time can be considered a universal cost for all the transport means. The results plotted in fig. \ref{fig5} highlight a collapse of the distributions for the bike and car mobility up to values $T/T_m\simeq 2.5 $, whereas the pedestrian distribution has a sharp peak at $T/T_m=0.5$. We remark that all the distributions have similar features with an initial fast increase for short trips and an exponential-like decaying, so that we consider the possibility that a single model could explain all the distributions pointing out the relevant parameters that characterize the use of different transport means.

\section{A survival model for human mobility}

Our point of view is that the statistical properties of human mobility can be described by a simple survival model based on the assumption of the existence of a mobility cost function: the travel time.  Our main assumption is that the perceived utility in the choice of different transport means is related to the travel time distribution for short trips, whereas the long trips distributions is justified by a Maximal Entropy Principle\cite{gallotti2012}.
We define $P(T)$ as the probability that a trip has a duration greater than $T$ and we propose a simple survival model to describe the empirical trip duration distributions assuming that the underlying microscopic dynamics can be described by a regular Markov process\cite{vankampen2007}
\begin{equation}
P(T+\Delta T)\simeq (1-\pi(T)\Delta T)P(T)
\label{markov}
\end{equation}
where $\pi(T)$ is the `stop' transition rate: i.e. the probability that a trip stops per unit time after a duration $T$. By definition the travel time distribution $p(T)$ is defined by
\begin{equation}
p(T)=-\frac{dP}{dT}
\label{distri}
\end{equation}
If $\lim_{T\to \infty}\pi(T)\to \beta$ we have an exponential decaying for $P(T)$. However, since the urban mobility is a consequence of the individual decision to perform certain activities, we model the transition rate $\pi(T)$ using the concept of \textit{utility function}. For a given transport means, let $U_c(T)$ and $U_s(T)$ a measure of the perceived utility to continue or to stop a trip of duration $T$ in the framework of a logit model: $U_c(T)$ could be related to the expected advantage get to perform the activity, and to availability of a mobility network for the chosen means, whereas $U_s(T)$ is the disadvantage to continue the trip after a duration $T$.
The simplest assumption is that $U_c(T)=U_c^0$ is constant to measure the initial utility to perform a trip by using a certain transport means, whereas $U_s(T)=\alpha T$ is an increasing function of $T$. Then we define $\pi(T)$ using the logistic function
\begin{equation*}
\pi(T)\propto \frac{e^{U_s(T)}}{e^{U_c(T)}+e^{U_s(T)}}
\end{equation*}
so that introducing the fatigue time scale $\beta^{-1}$ for the the proportionality factor, one gets
\begin{equation}
\pi(T)= \frac{\beta}{1+\exp(U_c^0-\alpha T)}= \frac{\beta}{1+\exp(-\alpha(T-T_c))}\qquad T_c=\frac{U_c^0}{\alpha}
\label{hazard}
\end{equation}
The duration $T_c$ can be interpreted as the characteristic duration associated to the convenience of using a certain transport means. In the continuous limit $\Delta T \to 0$ in eq. (\ref{markov}) we get the survival model
\begin{equation}
\frac{dP}{dT}=-\pi(T)P(T)
\label{survival}
\end{equation}
where $\pi(T)$ is the hazard function. The analytical solution of eq. (\ref{survival}) is explicitly written in the form
\begin{equation}
P(T)=e^{-\beta T}\left [\frac{1+\exp(\alpha T_c)}{1+\exp(-\alpha (T-T_c))}\right ]^{\beta/\alpha}
\label{survival_sol}
\end{equation}
so that we recover an exponential decay for $T\gg T_c$ and an inflection point when $\alpha T_c\gg 1$. Under this point of view the three parameters of the model (\ref{survival}) can be associated to physical observables for the understanding the statistical of human mobility. Simple algebraic manipulations provide the travel time distribution
\begin{equation}
p(T)\propto \frac{\exp(-\beta T)}{(1+\exp(-\alpha (T-T_c))^{\beta/\alpha+1}}
\label{distrit}
\end{equation}
so that when $T\gg T_c$ we have the exponential decaying with a characteristic time scale $\beta^{-1}$, whereas when $T\ll T_c$ if $\alpha$ increases $p(T)\to (\exp(-\alpha T_c))^{\beta/(\alpha+1)}$
The time $T_c$ is related to the distribution mode $T_\ast$ by
\begin{equation*}
T_\ast=T_c-\frac{1}{\alpha}\ln\frac{\beta}{\alpha}
\end{equation*}
The characteristic time scale $\beta^{-1}$ can be interpreted as a time cost for the chosen transport means (walking or bike-riding are really energy demanding activities, but also using car in traffic conditions is cause of fatigue) and trips durations $T\gg \beta^{-1}$ are very improbable. Similarly, if $\alpha T_c\gg 1$ is also improbable to observe travel times $T\ll \alpha^{-1}$ so that this time scale can be interpreted as a \textit{convenience time} scale to use that transport means.  We have the requirement $\beta/\alpha\ll 1$ since the convenience time scale should be shorter than the fatigue time scale. This implies that $T_c\simeq T_\ast$ defines the \textit{typical travel time} for the chosen transport means. Therefore the function (\ref{distrit}) does not only provide an interpolation of the empirical travel time distribution for pedestrian, bike and private car mobility, but it also allows to relate the parameter values to specific features of each transport means in order to understand the role of travel time as mobility energy. In particular, the comparison of the empirical parameters for the different transport means allows a better understanding of the relation between the statistical properties of urban mobility and a possible individual decision mechanisms in order to build microscopic models for multimodal mobility.

\section{Data analysis results}

\begin{figure}[tbp]
  \begin{center}
    \includegraphics[width=0.49\textwidth]{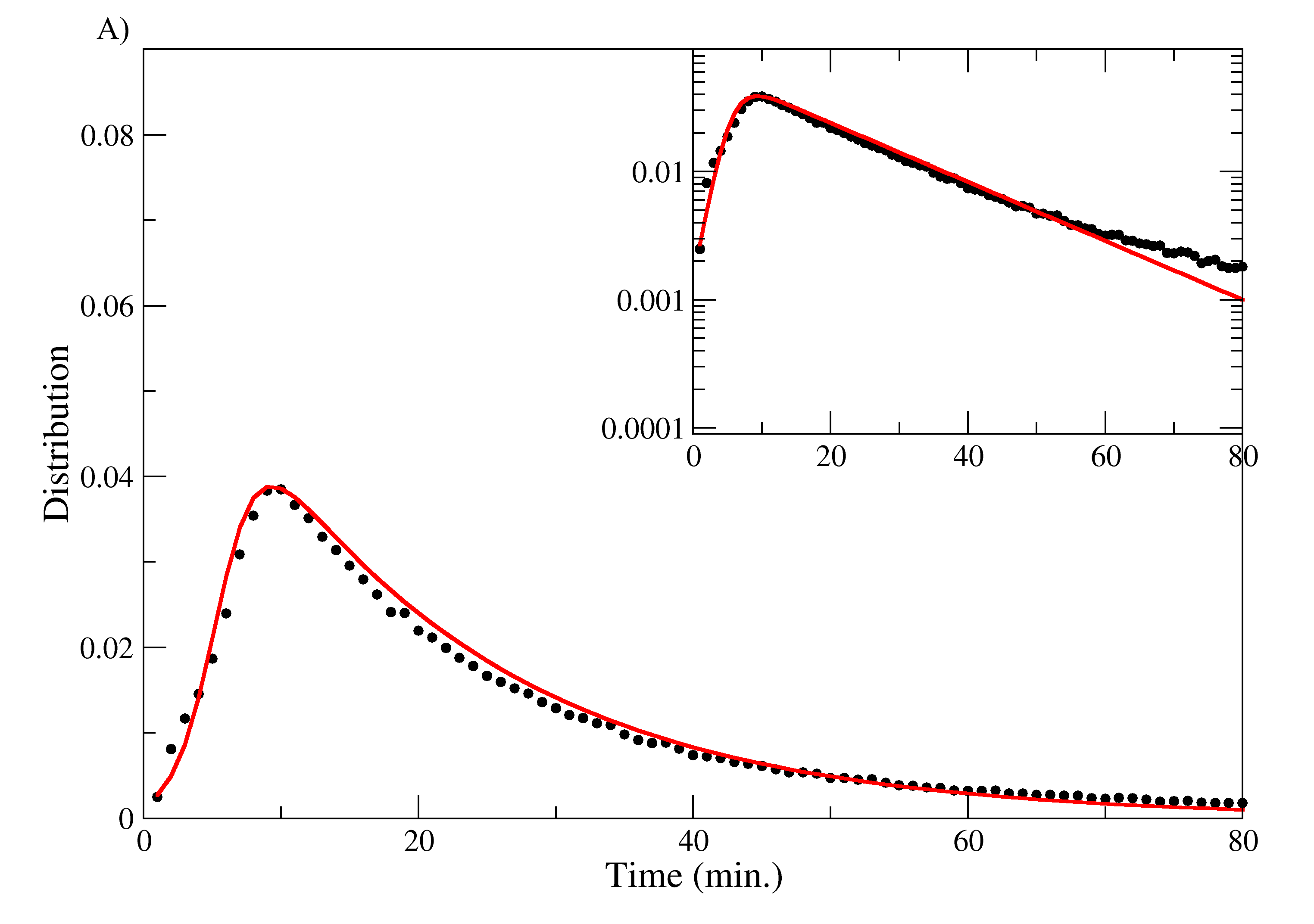}
    \includegraphics[width=0.49\textwidth]{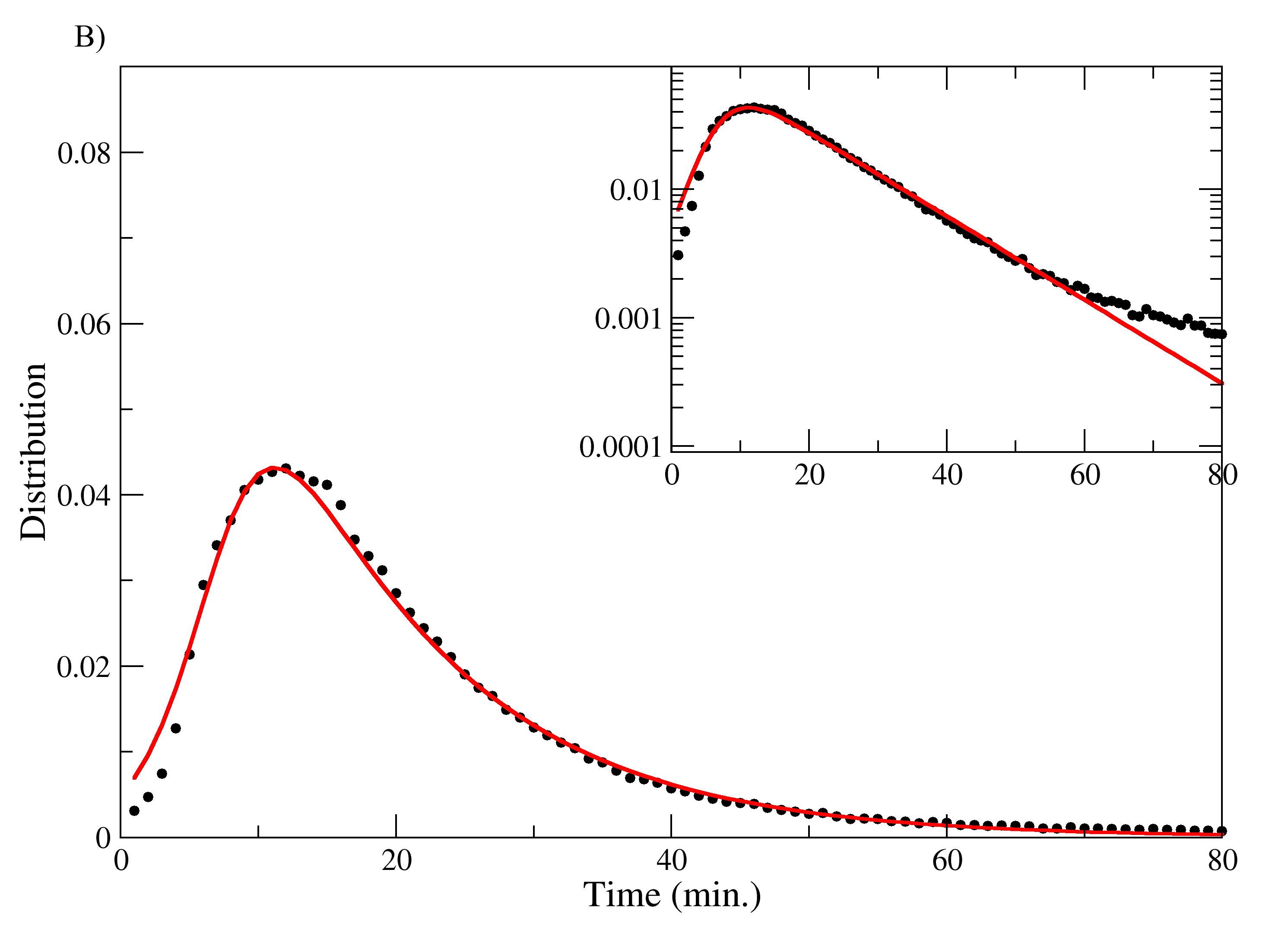}
  \end{center}
  \begin{center}
    \includegraphics[width=0.49\textwidth]{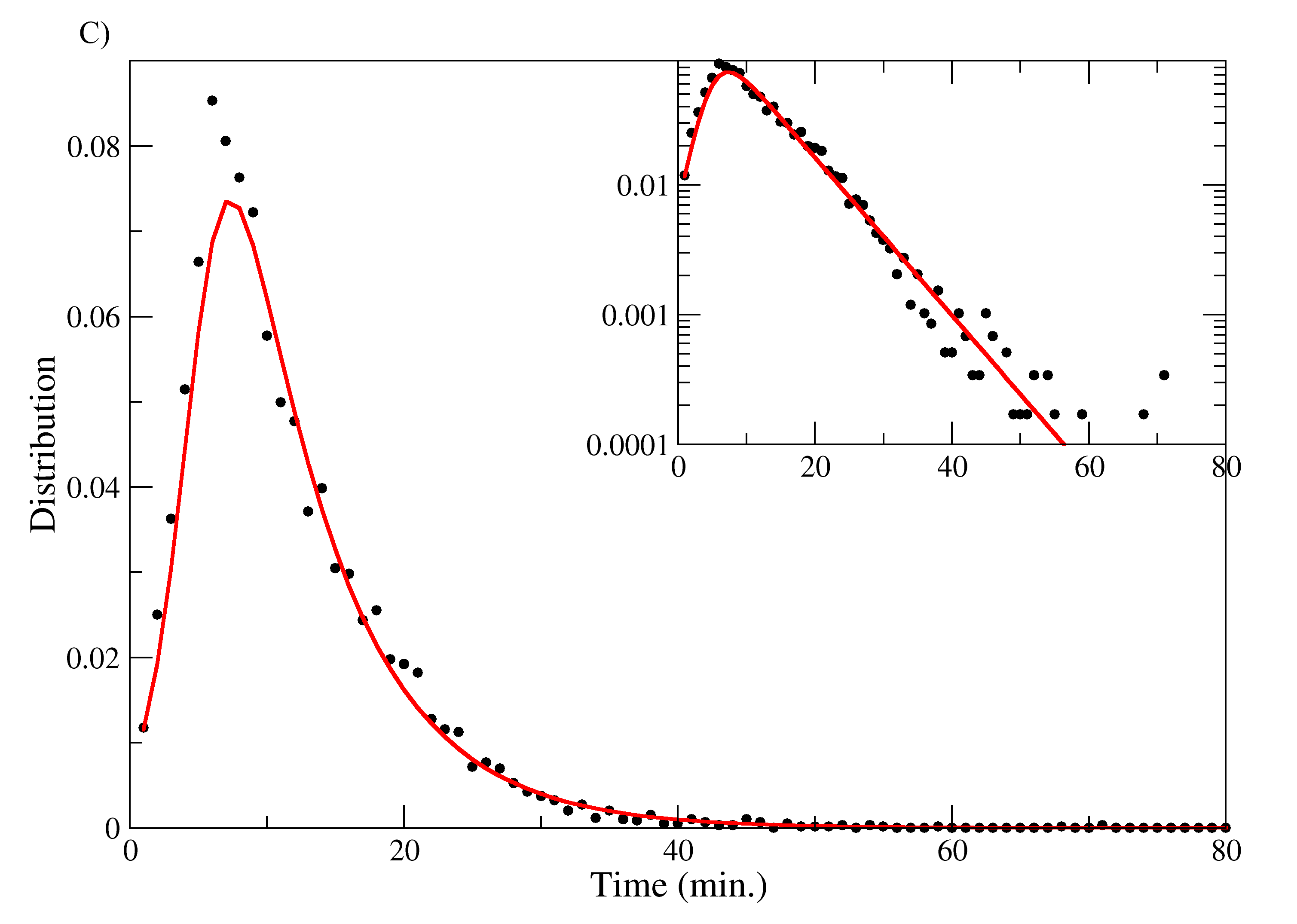}
    \includegraphics[width=0.49\textwidth]{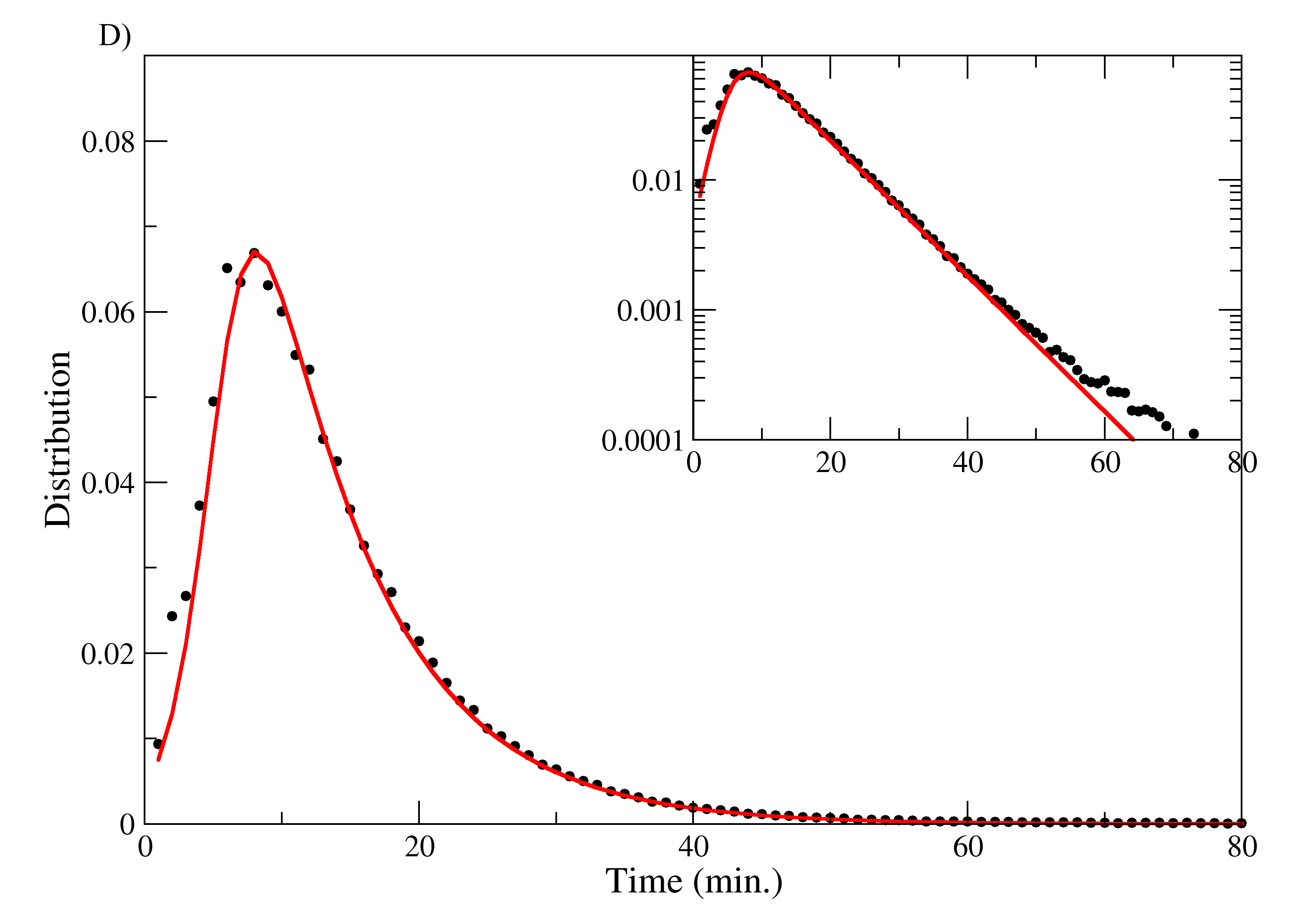}
  \end{center}
  \caption{In the figures A),B),C),D) we show the empirical travel time distributions (black dots) for the pedestrian, bike, car HC and car MA mobility with the computed interpolations (continuous lines) based on the survival model (\ref{survival}) with the parameter values reported in table \ref{table2}. The insets
  show the same plots in a semilog scale.}
  \label{fig6}
\end{figure}

\begin{figure}[tbp]
  \begin{center}
    \includegraphics[width=0.63\textwidth]{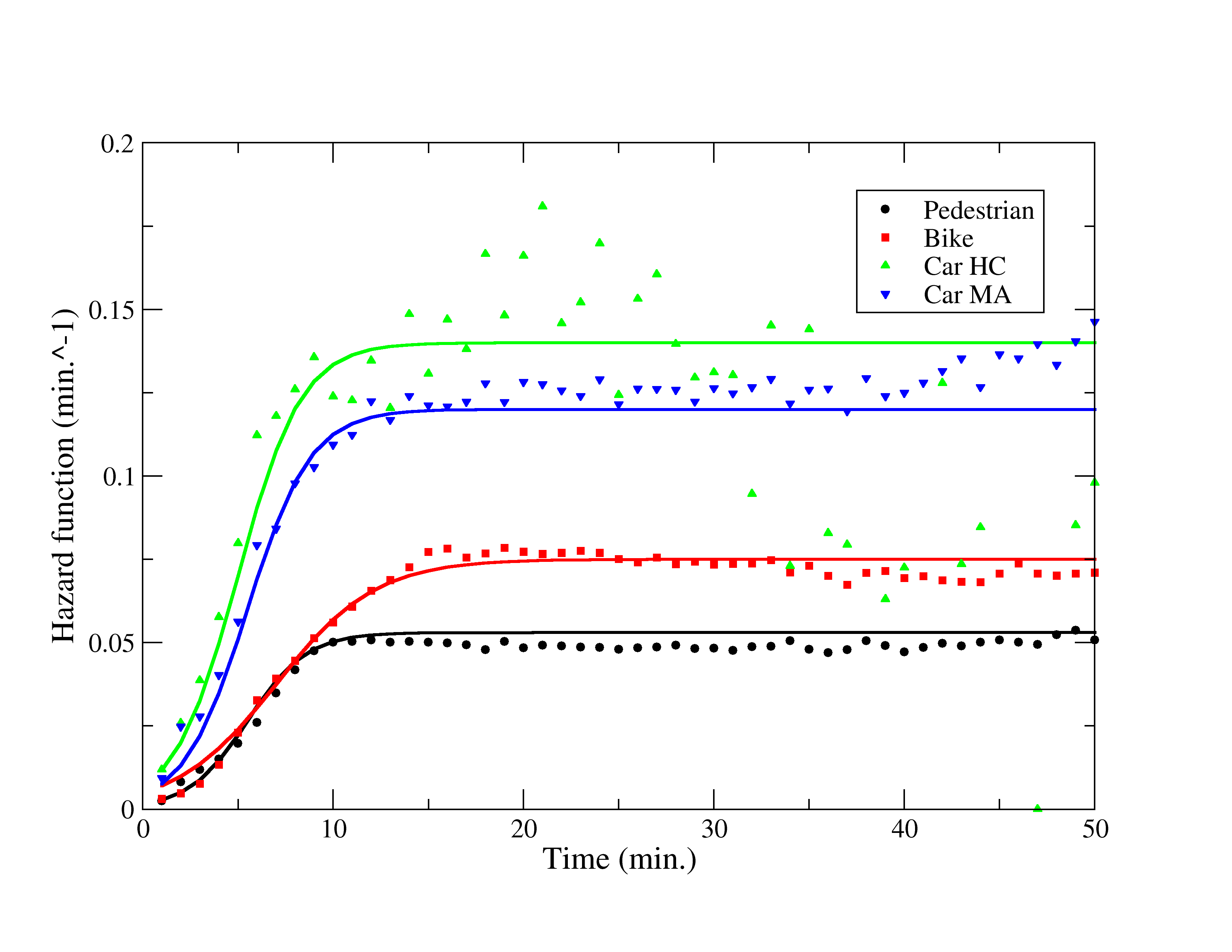}
    \includegraphics[width=0.63\textwidth]{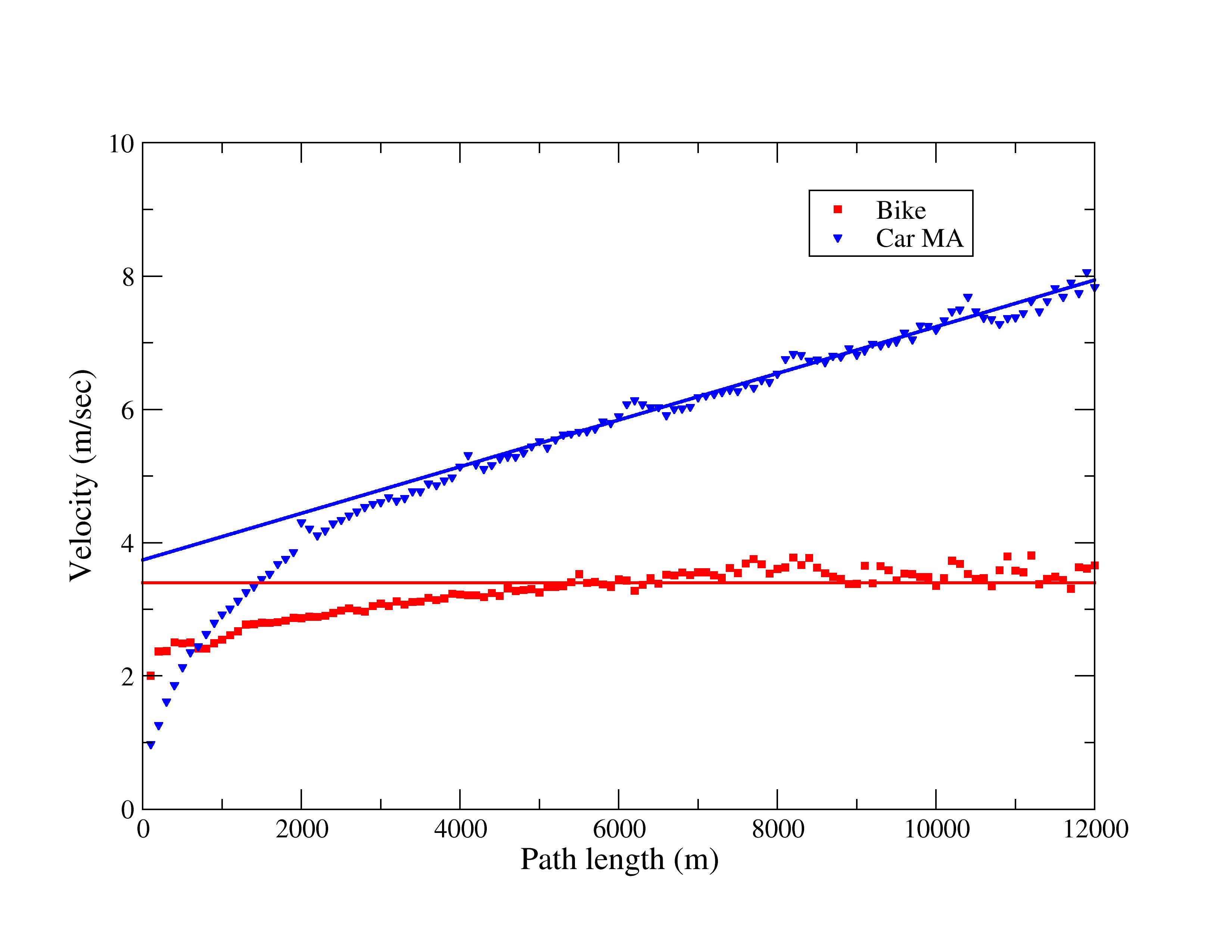}
  \end{center}
  \caption{Top picture: comparison between the empirical hazard functions $p(T)/P(T)$ (see eq. (\ref{survival})) for the different transport means with the theoretical hazard function (\ref{hazard}) using the parameters reported in table \ref{table2}. The fluctuations observed in the car HC points is due to the poor statistics when the travel time is long. Bottom picture: average velocity as a function of the trip lengths:
  the squares refer to the bike mobility whereas the triangle down to the car MA mobility. The straight lines are the result of the regression and show an asymptotic value of $3.4$ m/sec.
  for the bikes and a positive slope $3.5\times 10^{-4}$ sec.$^{-1}$ for the cars. The little bumps any 2 km are due to the spatial discretization of the long trajectories in the dataset
  and the  limited dimension of the considered area so that some trajectories are truncated.}
  \label{fig7}
\end{figure}

The interpolation procedure for the model (\ref{survival}) is performed first estimating the parameter $\beta$ from the exponential decaying and then computing the parameters of the hazard function by fixing $\alpha$ and $T_c$. Our aim is to show as the survival model is able to reproduce and to explain the main statistical features of the travel time distributions for different transport means. In table \ref{table2} we report the parameter values used for interpolating the empirical distributions and the corresponding plots are shown by fig. \ref{fig6}.
\begin{table}[h!]
  \begin{center}
    \begin{tabular}{l|c|c|r}
      \toprule
        transport means &  time cost $\beta^{-1}$ & convenience time $\alpha^{-1}$  & typical time $T_c$ \\
      \midrule
        pedestrians & 18.9 min.  &  1.5 min. & 5.5 min. \\
        bikes & 13.3 min.  & 2.6 min. &  7. min. \\
        cars HC & 7.1 min.   & 1.7 min.  &  5.0 min. \\
        cars MA & 8.3 min.  & 1.7 min. & 5.5 min. \\
      \bottomrule
    \end{tabular}
  \end{center}
  \caption{Interpolation parameters for the different travel time distributions: the statistical error on the parameter values is implied in the last digit.}
  \label{table2}
\end{table}
We remark that in all the considered cases the survival model provides a quite good interpolation of the empirical distributions and the main difference among the parameters is the time cost value which varies from $7.1$ min. for cars HC up to $18.9$ min. for pedestrians. More precisely, the time cost for car mobility is small ($7\div 8$ min. in both cases), which means that the private cars are perceived convenient since they have a smaller cost with respect to the bike and pedestrian mobility. It is noteworthy that convenience time scale $\alpha^{-1}=1.7$ min. is very short for the cars and the distribution mode is $T_\ast=9.5$ min. for cars HC and $T_\ast=11$ min. for cars MA (slightly greater than $T_c$ in both cases) suggesting the use of cars for very short trips. This is also in agreement with the path length distribution in fig. \ref{fig5}, where we do not obserse a clear drop down of the distribution at short path lengths for cars. This can be the consequence of a complex mobility performed by people for which the car allows to fulfill many daily activities, which individually require short trips.
We finally remark the similarity of the parameter values in the case of cars HC and cars MA even if the average velocities are very different in the two cases (see table \ref{table1}). This is consistent with the universality of the travel time cost for the urban mobility.
We have also compared the empirical hazard function computed as the ratio $p(T)/P(T)$ between the distribution function $p(T)$ and the stopping probability $P(T)$ with the theoretical hazard function $\pi(T)$ defined by eq. (\ref{hazard}). We remark as the hazard function (\ref{hazard}) describes correctly the behaviour of empirical curves for short travel times in all the considered cases.
The exponential character of the path lengths distribution of cars (see fig. \ref{fig5}) can be understood, since the convenience time scale is very short and the average velocity is much smaller for short trip than for long trips as shown in fig. \ref{fig7}. This apparent acceleration for long trips is the consequence of the non-homogeneity of the underlying road network that allows individuals to change their mobility strategy according to the trip length\cite{gallotti2016}.
We remark that this effect is not present for the bike mobility since the average velocity saturates at $V_m=3.5$ m/sec. (see fig. \ref{fig7}) and the road network can be considered homogeneous for the bike.
\par\noindent 
The travel time distribution for bikes has a longer convenience and fatigue time scales, so that the short trips are depressed in the travel time distribution. This fact could be interpreted since the use of bike is chosen when paths are no too short (otherwise one chooses to walk), and then people ride the bike for a relative long time (the mode of the distribution is $\simeq 17$ min.). The rescaling of the travel times by the means value $T_m$ (see fig. \ref{fig5} (bottom)) shows that the distributions for bikes and private cars tend to collapse suggesting that bikes and cars are used to perform a similar mobility with different time costs.
Finally, we remark that the walking mobility is characterized by a short convenience time scale but a long time cost (once one has decided to walk the time required to perform the mobility cannot be too little). The mode of the distribution is at $9.3$ min. less than the average value (see table \ref{table1}) since $T_c=5.5$ min. is small. The peculiarity of the pedestrian travel time distribution is the long time cost with short convenience and typical travel times, whose values are similar to that of the car mobility. Therefore people could choose to walking to perform with a wide range of travel time also for healthy implications of this activity.
As a matter of fact the distribution mode values are very similar for all the distributions. This could be another indication of the existence of a travel time budget related to people mobility in a city regardless the specific transport means.

\section{Conclusions}

The multimodality mobility is a key issue for the realization of a sustainable mobility in the future smart cities. The main goal of the paper is to show the existence of universal statistical properties of human mobility taking advantage from the availability of GPS data for single paths using different transport means. We have studied the mobility in the MA of Bologna where the app Bella Mossa collected data, for six months during 2017, on bike and pedestrian mobility and GPS data on private car trajectories were collected for insurance reasons (Octo Telematics dataset). Both the datasets may be affected by some bias since we have not a control on the statistical sample.
We have analyzed the distributions for the average velocities, path lengths and times for the pedestrian, bike and cars mobility distinguishing between the HC traffic and the mobilty in the whole MA. 
The average velocity distribution allowed to confirm the consistency of the BM dataset where the classification between pedestrian and bike mobility was given by the users and 
it pointed out as the high traffic affects the velocity distribution.
Our results suggest that the average value for the velocity distribution is the only relevant observable related to the traffic load at macroscopic level and that the shape of distribution should depend on the structure of the underlying road network. Indeed, the comparison of the normalized velocity distributions shows as the car mobility in the MA of Bologna has a different shape with respect to the pedestrian, bike and car HC distributions. A possible explanation is that the average velocity distributions reflect the heterogeneity of individual behaviours and structure of the road network. In the MA the road network has multilayer structure that allows the realization of complex mobility strategies. This is also confirmed by the dependence of the average velocity with respect to the path length in the MA (see fig. \ref{fig7}) and it is in accordance with previous results\cite{gallotti2016}. The path length distributions show by an exponential-like decaying for all the transport means that is consistent with a Maximal Entropy Principle and the existence of a finite mobility energy\cite{kolbl2003,gallotti2012}.
In particular, the car mobility contains very short path lengths (for car HC we estimate $L_m\simeq 2$ km) suggesting that the private cars are used to perform a mobility that cannot be reduced to an origin-destination mobility. Then we have considered the travel time distributions whose shape show universal features for the different transport means (see fig. \ref{fig5}) that could reflect both the existence of a mobility energy (i.e. the travel time budget concept\cite{marchetti1994}) and the individual choices of the transport means.
We addressed the problem of studying universal statistical properties of urban mobility by proposing a survival dynamical model based on three observables associated to three time scales: the time cost $\beta^{-1}$, the convenience time $\alpha^{-1}$ and the typical time $T_c$. If the time cost is directly related to a mobility energy, the convenience time and the typical time could be the effect of the individual decision process underlying the use of different transport means. Our results show that the survival model is able to reproduce the shape of the travel time distributions and to give an explanation for the observed collapse of the distributions for the bike mobility and the car mobility when we normalized the travel time by its average value.
The bike and car paths can realize the same mobility demand, with different time scales but a very similar spatial scale. The pedestrian mobility is an alternative mobility with a great time cost, but the energy required by the pedestrian mobility implies that the normalized travel time distribution has a different shape with a mode value different from the time cost.
The measure of the characteristic time scales for the different transport means in a city could be helpful for planning the future smart cities\cite{batty2012} to realize a sustainable mobility, and build microscopic models of urban mobility could integrate the survival model to introduce a decision mechanism that mimics the citizens choice of the transport means. Finally, specific governance policies (i.e. encouraging the use of electric bikes) could be developed to change the parameter values of the model as the time cost or the convenience time scale.

\section{Acknowledgments}
We thank SRM – Reti e Mobilit\'a S.r.l. for providing the Bella Mossa dataset for the bike and pedestrian mobility and Octo Telematics S.p.A. for providing the dataset for the private car mobility.

\section{References}

\appendix

\section{Datasets description} \label{app:data}

\begin{figure}[tbp]
  \begin{center}
    \includegraphics[width=0.7\textwidth]{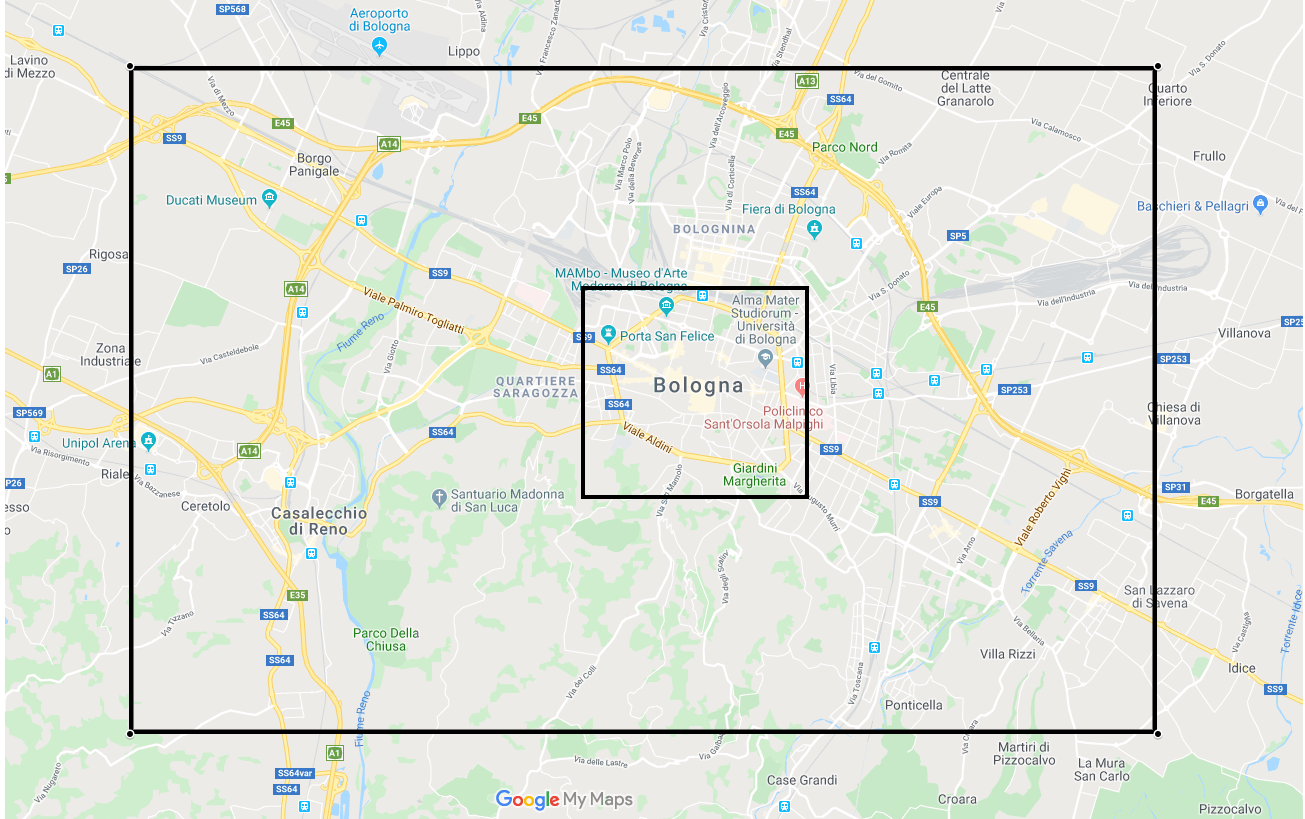}\vskip 1. truecm
    \includegraphics[width=0.7\textwidth]{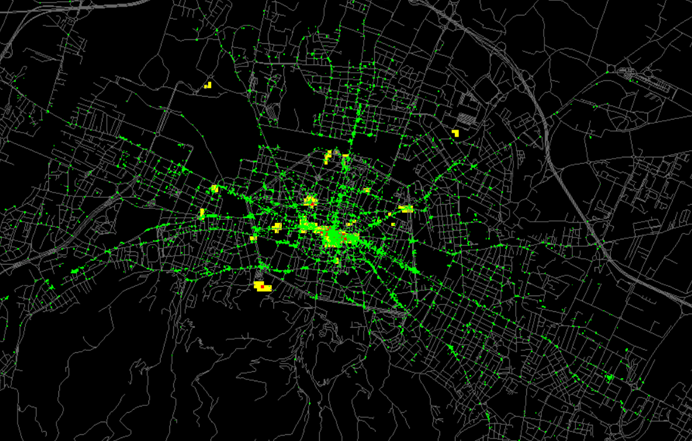}
  \end{center}
  \caption{Top picture: map of the Bologna metropolitan area: the larger rectangle encloses the whole area that has been considered by our mobility analysis, whereas the smaller one encloses the historical center of Bologna that is used to disaggregate the OT dataset. Bottom picture: distribution of the GPS points of the BM dataset whose trajectories have the endpoint in the Bologna historical center.}
  \label{fig1}
\end{figure}

The datasets used in the paper are not public and they have been made available thanks to non disclosure research agreements with 
SRM – Reti e Mobilit\'a S.r.l. for the BM dataset and Octo Telematics S.p.A. for the OT dataset on the private car mobility. Both the datasets contain GPS data on single paths
that are identified by an anonymous Id and in the Bella Mossa dataset one has information on the performed activity (walking or riding a bike) given by the users.
We have checked this information on the transport means by comparing the average velocity distribution associated to each trip with the expected typical velocities of a pedestrian or of a bike.
In both cases the recorded paths were performed in the metropolitan area of Bologna that is defined in the fig. \ref{fig1} (top). But we separate the mobility in historical center area in the analysis since it is the main attraction area for an origin destination mobility and many restriction policies have been applied to reduce the traffic load in that area.
To show the quality of the datasets we plot in fig. \ref{fig1} (bottom) the distribution of the initial points of the trajectories whose end point is in the historical center: the point distribution highlights the attractiveness of the HC for slow mobility. 
The BM dataset contains data collected by using a specific app offered to the citizens with the aim to improve the sustainable mobility:
when activated, the app allowed to collect GPS data each 2 sec. on the trajectory performed by an individual, that has also the possibility of declaring the transportation mean associated to each trip. The data recording starts when the app is switched on and terminates when the app is switched off and each trajectory is associated to an anonymized Id so that there is not the possibility of tracking individuals. The BM initiative has involved $\simeq 10^4$ citizens in Bologna with an average number of $\simeq 4000$ trajectories recorded each day and a total of \numprint{136000} bike trajectories and \numprint{310000} pedestrian trajectories after a filtering procedure on the very short trips. Even if we have not a control on the statistical sample, the number of recorded trajectories is sufficiently big to highlight the statistical properties of bike and pedestrian mobility in Bologna.\par
\begin{figure}[tbp]
  \begin{center}
    \includegraphics[width=0.72\textwidth]{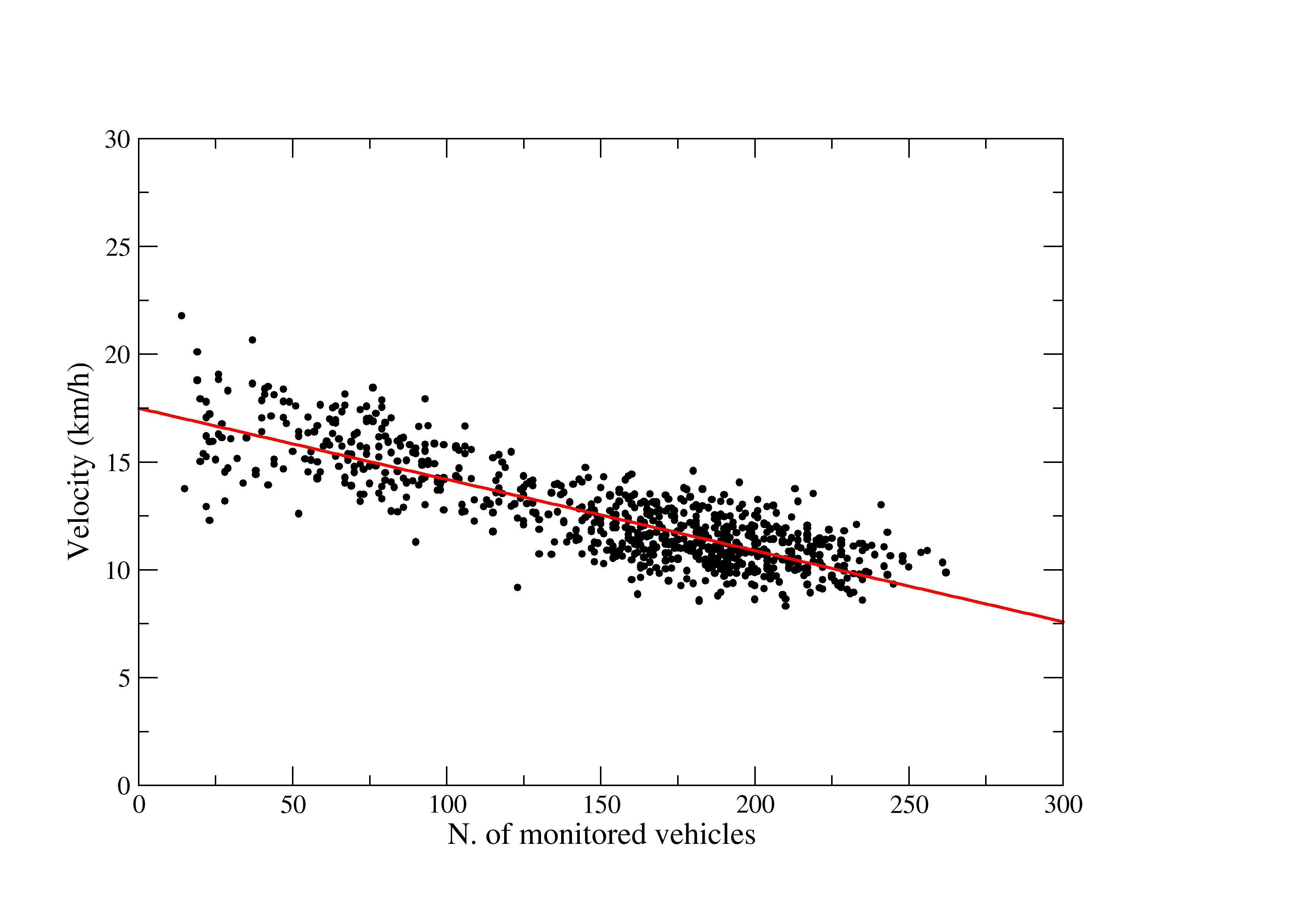}
    \includegraphics[width=0.72\textwidth]{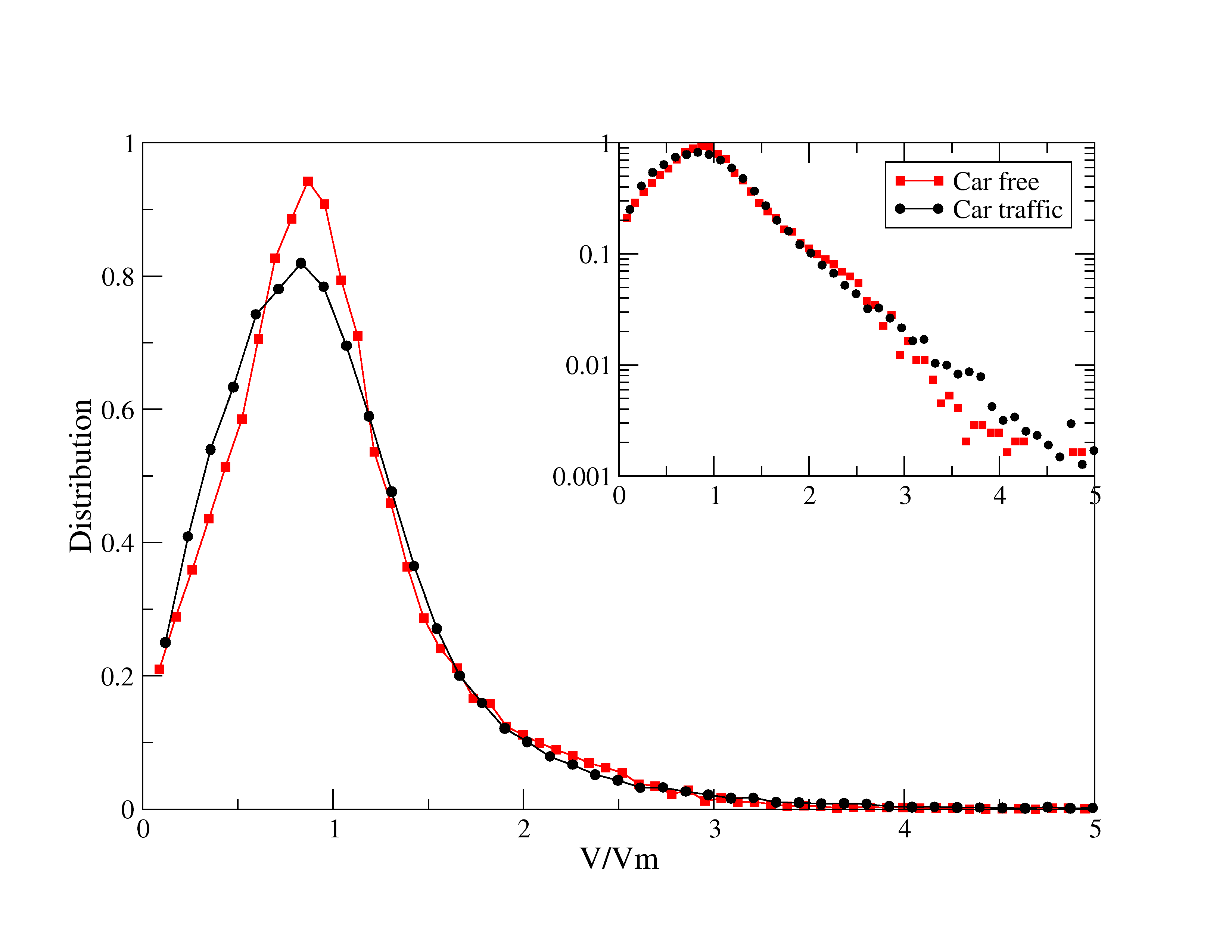}    
  \end{center}
  \caption{Top picture: fundamental diagram average-velocity versus traffic load (number of monitored vehicles present) computed using the car trips of the OT dataset performed in the historical
center during September 2016; each point corresponds to a time interval of 30 minutes and the right line is the result of a linear regression. Bottom picture: distribution of the normalized average velocity $V/V_m$ where $V_m$ is the mean value for the vehicle trips: the circles refer to rush hours 7:30-8:30 and the squares to a free traffic regime 22:30-23:30 in the whole Bologna MA. The inset shows the same
distributions in a semilog scale suggesting an exponential decaying for large velocities.}
  \label{fig3a}
\end{figure}

\noindent
The OT dataset contains information on the private car mobility in the MA of Bologna that are recorded for insurance reasons. The dataset contains a sampling of the vehicle trajectories at a spatial scale of 2 km or at a time scale of 30 sec, with the quality of a GPS datum. The data refer to the month of September 2016 and contain $\simeq 6\times 10^5$ trajectories on a sample of private vehicles moving
in the area. Previous studies\cite{bazzani2010,gallotti2012} have pointed out that the sample penetration of this dataset can be estimated as $\simeq 5\%$ of the whole daily vehicle population, by comparing the expected traffic flow along the main roads with the traffic flow data recorded by magnetic coils. 
In the case of urban mobility the reconstruction of the real trajectories is a difficult task due to the complexity of the urban road network and the spatial scale of the sampling, however we have information on the location of the starting and ending points of each trip (that corresponds to the power on and off of the engine). Moreover the duration and the path length are automatically computed by the devices installed on the vehicle. Therefore, we have good quality data on the path lengths, the duration and the average velocity of private urban mobility with the possibility of distinguishing rush hours from normal traffic conditions. To avoid an overcounting of the short trips (one could switched off the engine during short term stops) we apply an algorithm that joins the trips when the stopping time is less than 1 minutes and there is a continuity in the direction of the successive trip.
To check the quality of the dataset we computed the fundamental diagram for the car trips recorded in the HC area of Bologna to dtect the effect of traffic load on the average velocity. We have restricted the analysis to the trips inside the HC during the working days to consider the traffic dynamics on a homogeneous road network as suggested in \cite{daganzo}, and we also assumed that the traffic load in Bologna is directly proportional to the number of monitored vehicles present in the considered time interval. Each point is computed dividing the total traveled length by the total travel time of the moving vehicles during the considered time interval of $30$ minutes. This analysis provides a congestion measure for the whole road network that cannot be related to the average velocity of individual trips. The fundamental diagram (see fig. \ref{fig3a} (top)) highlights the existence of different traffic regimes in the urban road network according to the traffic load that can be distinguished by a different average velocity. 
In the fundamental diagram we observe as the points cluster along a straight line with a negative slope that represents the expected decrease of the average velocity when the traffic load increases. The regression straight line suggests a free traffic average velocity $\simeq 17.5$~km/h for the HC road network, that reduces down to $10$~km/h in case of traffic. Moreover, the points corresponding to high traffic regime of the city do not follow the regression line and suggest the existence of a limit velocity for the traffic regime of $\simeq 10$~km/h. 
\par\noindent
If one disaggregates the data according to the traffic load in the whole area, the corresponding velocity distributions are characterized by different average values: $V_m=4.20$~m/sec during rush hours and $V_m=5.76$~m/sec for free traffic. However, the distributions for the normalized velocity $V/V_m$ are very similar in both cases (see fig. \ref{fig3a} (bottom)) suggesting that the average velocity is directly related to the traffic load and the possible congestion effects increase slightly the distribution values corresponding to the low velocities during rush hours. On the contrary, in the case of free traffic, the distribution is more peaked around the average value $V/V_m=1$. It is also interesting to observe the exponential decaying for large values of the normalized velocities that is the same in both case. By definition, the average velocity for a given path length $L$ (not too small) is $V=L/T$ where $T$ is the total travel time that depends on the traffic condition on the road network (we recall that both $L$ and $T$ are directly measures
by the GPS devices in the datasets). The average value $T_m$ for the travel time is related to the traffic load on the road network and it can be used to measure the congestion degree. Assuming that the decreasing of the travel time by a quantity $\Delta T$ with respect to an expected value is modeled by a Poisson random process (this model could also simulates the heterogeneity of individual behaviour), one has the probability distribution
\begin{equation}
P(\Delta T)\propto p^{-\Delta T/T_m}\qquad \Delta T\ge 0
\label{distritime}
\end{equation}
where $p \in [0,1]$ is a suitable constant that defines the probability of the relative fluctuations. Then the corresponding relative velocity increase is
\begin{equation*}
\frac{\Delta V}{V_m}\simeq \frac{\Delta T}{T_m}
\end{equation*}
where $V_m=L/T_m$ so that, if the travel time fluctuations do not depend on $L$, one gets an exponential decaying in the probability distribution of the normalized velocity $V/V_m$ independently from the traffic condition, which essentially modifies $V_m$. This remark is also consistent with the normalized average velocity distributions plotted in the fig. \ref{fig3} of the main text.
The distributions suggest as the probability $p$ in eq. (\ref{distritime}) could depend on the features of the mobility network as well as on the heterogeneity of individual behaviors. According to this assumption,
we remark that the pedestrian and bike mobility share the same road network with the car mobility in the HC. On the contrary, the
country roads network and the highway increase the probability $p$ in the MA since the road network can
be viewed as a multilayer road network with different travel velocities for the different layers and the individuals have the possibility to improve
the mobility efficiency reducing the travel time (therefore increasing $p$) performing long trips.

\par\noindent

\end{document}